\documentclass[twocolumn,showpacs,preprintnumbers,amsmath,amssymb]{revtex4}

\usepackage{graphicx}
\usepackage{dcolumn}

\begin{document}

\title{Longitudinal Zeeman slowers based on permanent magnetic dipoles}

\author{Yuri B. Ovchinnikov\thanks{\email{yuri.ovchinnikov@npl.co.uk}}}

\affiliation{National Physical Laboratory, Hampton Road, Teddington,
Middlesex TW11 0LW, United Kingdom}

\begin{abstract}
Longitudinal Zeeman slowers composed of arrays of compact discrete neodymium magnets are proposed. The general properties of these slowers, as well as specific designs of short spin-flip Zeeman slowers for Sr and Rb atoms are described.   
The advantages of these slowers are their simplicity, low cost and absence of consumed electrical power and corresponding water cooling.
The smoothness of the magnetic field together with ease of adjustability makes it possible to operate 
these slowers near the theoretical limits of deceleration, making them more compact and efficient.
\end{abstract}


\maketitle

\section{Introduction}
\label{intro} 

A Zeeman slower is a special magnet, which allows continuous laser 
cooling of thermal atomic beams with just a single laser beam of fixed frequency. 
The purpose of this magnet is to compensate the Doppler frequency shift
of the cooling atomic transition by the spatially-dependent Zeeman frequency shift, 
which provides a continuous resonant interaction between the cooling laser light and the decelerated atoms 
for all atomic velocities within the capture range of the slower.
As the most intense sources of cold atoms, ovens combined with Zeeman slowers are widely used in different atomic physics experiments,
including producing Bose Einstein Condensates \cite{BEC,Streed} and optical atomic clocks \cite{Takamoto}, 
which demand a large number of cold atoms.

A standard longitudinal Zeeman slower \cite{Phillips} uses a special current-carrying coil to create a 
proper spatial distribution of the magnetic field along its axis.

In our previous work \cite{Ovchinnikov1}, Zeeman slowers based on arrays of permanent magnetic dipoles (MD) were proposed. The first of these spin-flip transverse Zeeman slower for Sr atoms \cite{Ovchinnikov2}, based on compact
neodymium permanent magnets, was recently implemented in the NPL Sr optical clock and has already demonstrated its efficiency, convenience and reliability.
In spite of the simplicity of the transverse MD-Zeeman slowers (where the magnetic field is directed transversely to the direction of the atomic and cooling laser beams) such slowers have some limitations.
First, the transverse Zeeman slower demands twice as much laser power, compared to standard longitudinal slowers.
For linear polarization of the cooling light, only one of its two circular-polarization components of different helicity is in resonance with the corresponding Zeeman shifted transition of the slowed atoms.
Second, the spin-flip (the slower, in which the magnetic field is changing its sign) transverse Zeeman slower is working only for atoms without 
(or with small) Zeeman splittings of the ground internal state, like Sr or Yb, for example.
For atoms with essential splitting of the ground state, like Li, Na, Rb, Cs, Cr the linear polarization of the cooling light of the transverse Zeeman slower prevents it from optically pumping atoms to the correct (slowed) magnetic substate in the region where the the magnetic field turns to zero, which leads to loss of the decelerated atoms at that region. This last fact has been confirmed by the recent tests of the transverse MD-Zeeman slower for Li and Cr atoms \cite{McL}.

Recently a longitudinal Zeeman slower for Rb atoms, based on an 8-bars Halbach configuration of permanent magnets, has been successfully demonstrated \cite{Cheiney}.

This paper considers longitudinal spin-flip Zeeman slowers, which are using just two or four rows of standard cylindrical neodymium magnets, configured according to the original general proposal \cite{Ovchinnikov1}. The simplest of these Sr longitudinal slowers has already been successfully tested on the NPL optical clock setup \cite{Hill}.

In the first chapter of this paper, a short revision of the general theory of Zeeman slowers is given.
The second chapter describes a general approach to the design of longitudinal MD-Zeeman slowers based on arrays of magnetic dipoles (MD) as well as the two designs of slower for Sr atoms.
The fourth chapter gives an example of a design of a spin-flip longitudinal slower for Rb atoms.   
Finally, in the conclusion the main results are summarized.

\section{General theory of a Zeeman slower}
\label{sec:1}

A Zeeman slower decelerates atoms by resonant scattering
of photons from a counterpropagating laser beam.
It is supposed that the monochromatic laser light of frequency $\omega$ resonantly interacts with some 
isolated and closed atomic transition of frequency $\omega_0$, such that the corresponding interaction
can be described with a two-level model of an atom. In addition, it is assumed that the atom is 
placed in the magnetic field of a Zeeman magnet, which is characterized by a spatially dependent 
amplitude $B(z)$. We assume also that this magnetic field leads to a Zeeman shift of the frequency of the atomic 
transition of $\mu' B(z)/\hbar$, where $\mu'$ is the magnetic moment for the atomic transition.   
The average spontaneous light pressure force on the atoms in the direction of propagation of 
the laser beam, given by
\begin{equation}
F(v,z)=\frac{\hbar k \Gamma}{2} \frac{s_0(z)}{1+s_0(z)+4[\delta_0+kv+\mu' B(z)/\hbar]^2/\Gamma^2},
\end{equation}
where $k$ is the wave vector of light, $s_0(z)$ is the local
on-resonance saturation parameter of the atomic transition,
$\Gamma$ is the linewidth of the transition, $\delta_0=\omega-\omega_0$ is the
laser frequency detuning and $v$ is the velocity of the atom. 
The velocity dependence of the force is determined by the effective frequency detuning
$\Delta_{eff}=\delta_0+kv+\mu' B(z)/\hbar$,
which includes the Doppler shift $kv$ of the atomic frequency. 
The dependence of the relative magnitude of the force on the effective frequency detuning $\Delta_{eff}$, 
for $s_0(z)=2$, is shown in fig.\,1.

\begin{figure}
\includegraphics[scale=0.45]{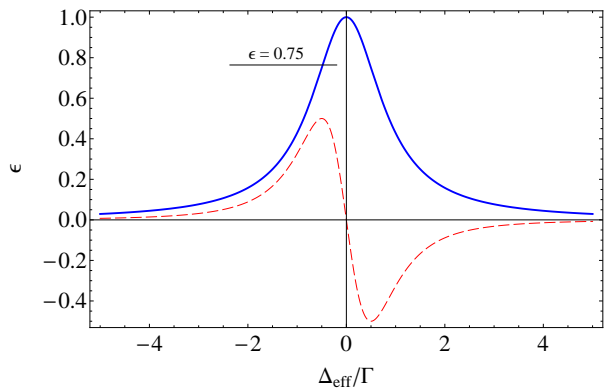}
\caption{The solid line gives the dependence of the relative light force, characterized by the coefficient $\epsilon$ (see the text), on the relative effective frequency detuning of the cooing light $\Delta_{eff}/\Gamma$. The dashed line gives the same dependence of the derivative of the force.}
\end{figure}

The maximum value of the local deceleration, provided by the force, is
achieved at exact resonance, $\Delta_{eff}=0$, and is given by
\begin{equation}
a_{max}(z)=\frac{\hbar k \Gamma}{2 m} \frac{s_0(z)}{1+s_0(z)},
\end{equation}
where $m$ is the mass of the atom. Although the maximum
deceleration can be used to estimate the shortest possible
length of a Zeeman slower, this can not be realized in practice. At
exact resonance an equilibrium between the inertial force of the
decelerated atoms and the light force is unstable and any
slight increase of the atomic velocity due to
imperfection of the magnetic field distribution or spontaneous
heating of atoms will lead to a decrease of the decelerating force
and subsequent loss of atoms from the deceleration process. In
practice the deceleration of atoms is realized at a fraction of
the maximum deceleration
\begin{equation}
a(z)=\epsilon a_{max}(z),
\end{equation}
where $\epsilon<1$ \cite{Ovchinnikov1}. Note that the coefficient $\epsilon$ in our
case corresponds to the ratio between the reduced local acceleration
and the maximum possible acceleration at the same location, which
is a function of the local saturation parameter $s_0(z)$. In a standard
treatment \cite{Napolitano,Metcalf} a similar coefficient
$\eta$ appears, which relates the actual deceleration to the maximum possible
deceleration at infinite intensity of laser light. 
The two coefficients $\epsilon$ and $\eta$ are related as $\epsilon=\eta s_0(z)/(1+s_0(z))$.
Therefore, the constant coefficient $\epsilon$ corresponds to a variable coefficient $\eta$, 
the value of which depends on the local intensity of light.

When the local deceleration $a(z)$ is less than $a_{max}$, which corresponds 
to $\epsilon<1$ the decelerated atoms stay on the low-velocity wing
of the Lorentzian velocity profile of the light force (1) and undergo
stable deceleration (see fig.\,1). The corresponding frequency detuning 
of the cooling light from the optical transition of the decelerated atoms at the active part of a 
Zeeman slower is determined by

\begin{equation}
\Delta_{eff}^*(z)=-\frac{\Gamma}{2} \sqrt{(1+s_0(z))\frac{1-\epsilon}{\epsilon}}.
\end{equation}

In a general case when the saturation parameter $s_0(z)$ changes due to convergence of the cooling laser beam and its absorption by decelerated atoms, the frequency detuning $\Delta_{eff}^*$ also essentially changes along the axis of the slower.   

The optimal offset of the equilibrium velocity $v(z)$ from
the resonant velocity is achieved at a point where
the derivative of the force (1) reaches its maximum, because the
damping of the relative motion of atoms around this point is
maximal. It is easy to show that within our definition of the
$\epsilon$ coefficient (3) this optimal cooling condition is
achieved exactly at $\epsilon=0.75$ 
at each local intensity of light inside the slower.  
The damping coefficient of the force, which is maximal at $\epsilon=0.75$, 
reaches its maximum at $s_0=2$ and slowly decreases with further increase of the intensity of the laser
field.

The calculation of the required spatial distribution of the magnetic field in a Zeeman slower
proceeds as follows. First, the actual velocity
of the slowing atoms is calculated numerically according to the formula

\begin{equation}
\frac{dv(z)}{dz}=\epsilon \frac{\hbar k \Gamma}{2 m} \frac{s_0(z)}{1+s_0(z)} \frac{1}{v(z)},
\end{equation}

With this approach it is easy to include the spatial variation of the saturation parameter $s_0(z)$ due to convergence
of the cooling laser beam and even due to absorption of its light by the decelerated atoms \cite{Ovchinnikov1}.
Based on the calculated slowing velocity $v_z(z)$,  the corresponding distribution of the magnetic field can be derived from the 
equality $\Delta_{eff}(z)=\Delta_{eff}^*(z)$ as
\begin{equation}
B(z)=\hbar \left(-\delta_0-kv(z)+\Delta_{eff}^*(z)\right)/\mu'.
\end{equation}

This equation accepts multiple solutions, which correspond to different constant offcet of the magnetic field and corresponding
frequency detuning $\delta_0=\Delta_{eff}^*(z_f)-kv(z_f)-\mu'B(z_f)/\hbar$ of the slowing light, where $z_f$ is the coordinate of 
the maximum of the magnetic field at the end of the slower. In this paper we are considering symmetric spin-flip Zeeman slowers, in which 
the magnitudes of the magnetic field at both ends of the slowers are equal and of opposite sign.

\section{Longitudinal MD-Zeeman slowers for Sr atoms}
\label{sec:2}

\subsection{General approach to longitudinal MD-Zeeman slowers}

The field of a single magnetic dipole placed at the origin of a Cartesian system of
coordinates and oriented along the z-axis is
described by the formulae
\begin{eqnarray}
B_x=\frac{\mu_0 M}{4 \pi} \left[ \frac{3xz}{r^5} \right] \nonumber \\
B_y=\frac{\mu_0 M}{4 \pi} \left[ \frac{3yz}{r^5} \right] \nonumber \\
B_z=\frac{\mu_0 M}{4 \pi} \left[ \frac{2z^2-x^2-y^2}{r^5} \right],
\end{eqnarray}
where $r=\sqrt{x^2+y^2+z^2}$ and $M$ is the magnetic moment of the dipole.

In this paper the longitudinal Zeeman slowers based on arrays of magnetic dipoles, oriented parallel 
to the $z$-axis of the slower, are considered.

The spatial distribution of the magnetic field generated by an array of such dipoles is 
determined by summation of the corresponding field amplitudes, produced by each of the individual dipoles. 
An array of magnetic dipoles, oriented parallel to the z-axis and 
placed in the X0Z-plane with coordinates of individual dipoles of $x=x_i,y=0,z=z_i$, are generating
the magnetic field with distribution of the $B_z$-component at the z-axis of

\begin{equation}
B_z=\frac{\mu_0 M}{4 \pi} \sum_{i=1}^N  \left[ \frac{2(z-z_i)^2-(x-x_i)^2-y^2}{((x-x_i)^2+y^2+(z-z_i)^2)^{5/2}} \right].
\end{equation}

For a symmetric distribution of the magnetic dipoles around the
$z$-axis, all other components of the magnetic field at this axis are equal to zero.
For positions away from the $z$-axis, the magnitudes of the transverse components of the field, $B_x$ and $B_y$, can be computed in a similar way.

\begin{figure}

\includegraphics[scale=0.4]{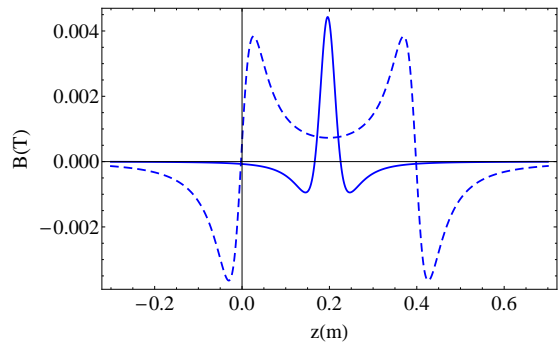}
\caption{Spatial distribution of $B_z$ component of the magnetic field along the $z$-axis, generated by two continuous line-like arrays of magnetic dipoles, oriented parallel to that axis and located at the same distance $R=4$\,cm from it. The solid line corresponds to the short array, $L_{short}=2$\,cm, and the dashed one to a long array of length $L_{long}=40$\,cm.}
\end{figure}

The solid line in fig.\,2 shows the spatial distribution of the $B_z(z)$ component of the magnetic field along the $z$-axis for a line-like continuous array of magnetic dipoles covering a length $L_{short}=2$\,cm with total magnetic moment $M=3.1$\,A$\cdot$m$^2$, oriented parallel to this axis and placed at a distance of $x=R=4.0$\,cm from it. The dashed curve in fig.\,2 shows the spatial distribution of the field at the $z$-axis generated by a $L_{long=}40$\,cm long continuous line-like array of magnetic dipoles with total magnetic moment $M=62.0$\,A$\cdot$m$^2$, which is located at the same distance of $x=R=4.0$\,cm from the $z$-axis. It means, that the linear density of the magnetic moments of these two arrays is the same. One can see that, in spite of the much larger total magnetic moment of the longer array of dipoles, its field amplitude is smaller than the field amplitude of the short array of magnetic dipoles. Moreover, for an infinitely long array of longitudinal magnetic dipoles, the amplitude of its outside field becomes equal to zero. It is easy to show that the $B_z$-component of the magnetic field of such an array is maximal when its length is $L\simeq0.4R$, while its spatial distribution along the $z$-axis is broadened but still has a single maximum. 

\begin{figure}
\includegraphics[scale=0.4]{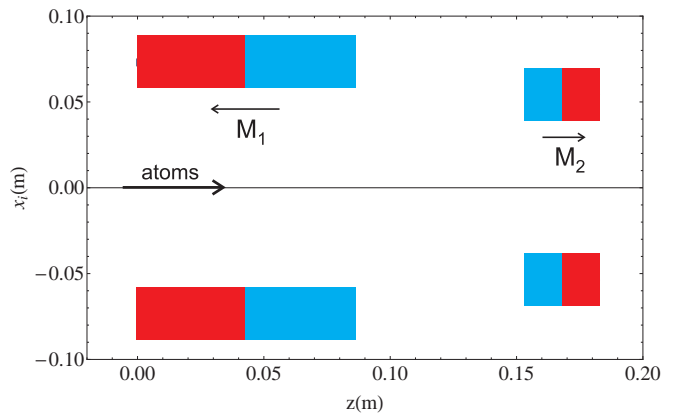}
\caption{Simplest longitudinal spin-flip MD Zeeman slower for Sr atoms.}
\end{figure}

For laser cooling of Sr atoms, the $^1S_0 \rightarrow ^1P_1$ transition with wavelength $\lambda=461\,$nm and
natural linewidth $\Gamma=2 \pi \times 32\,$MHz is used.
one of the simplest longitudinal spin-flip Zeeman slowers for Sr atoms, based on magnetic dipoles, is shown in fig.\,3.
It consists of four cylindrical neodymium magnets, placed symmetrically with respect to the $z$-axis of the slower in the X0Z plane. The magnetic moments of the long and short magnets are directed parallel to the $z$-axis with opposite polarity. All the magnets have a diameter of $30$\,mm. The base of the two long magnets is located at $z=0$ and $x=\pm74$\,mm and their length is equal to $88$\,mm.
The length of the short magnets is equal to $32$\,mm, while there distances from the axis of the slower are $x=\pm54$\,mm.
The gap between the magnets is equal to $65$\,mm. Therefore, the smaller end magnets are located closer to the axis of the slower, which provides a larger gradient of the magnetic field at the end of the slower.
The total magnetic moment of each of the long magnets is $M_1=B_rV/\mu_0=54.45$\,A$\cdot$m$^2$, where $V$ is the volume of the magnet, $B_r=1.1$\,T is the residual induction of a standard neodymium magnet and $\mu_0=4 \pi\cdot10^{-7}$H$\cdot$m$^{-1}$ is the permeability of free space. The corresponding total magnetic moment of the short magnets is equal to $M_2=19.8$\,A$\cdot$m$^2$.

\begin{figure}
\includegraphics[scale=0.4]{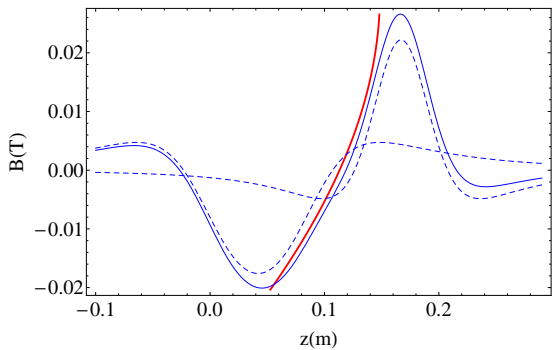}
\caption{The optimal magnetic field distribution (solid line) and the measured
field of the transverse MD Zeeman slower.}
\end{figure}

The central solid line in fig.\,4 shows the target spatial distribution of the magnetic field, which was calculated for $\epsilon=0.9$, capture velocity $v_c=330$\,m/s, final velocity $v_f=25$\,m/s, slowing length of about $12$\,cm, a saturation parameter of the cooling laser beam at the end of the slower of $s_0=1.0$ and a convergence angle of the laser beam, $\theta=0.012$. The design frequency detuning of the cooling laser beam is $\delta_0/2\pi=434$\,MHz. The slower is designed for $\epsilon=0.9$ to provide maximum deceleration of atoms, which minimises the length of the slower and decreases the transverse broadening of the atomic beam due to its heating by spontaneously scattered photons. It is worth mentioning that the operation of a Zeeman slower at almost critical deceleration ($\epsilon=0.9$) is possible only when its spetial distribution of the magnetic field is smooth enough, as it is in the MD Zeeman slowers considered here. The thin solid line shows the magnetic field distribution along the axis of the slower, which was calculated with a MD model (8), where each magnet was substituted with a row of point-like magnetic dipoles, separated from each other by a distance of 4\,mm, with the same total magnetic moment as the modeled magnet. The lines are slightly shifted with respect to each other along the $z$-axis solely to see better the agreement between their slopes.
The two dashed lines correspond to the spatial magnetic field distributions produced by each (left and right) half of the slower. One can see that the $z$-positions of maximum and minimum of the magnetic field are located nearly exactly where the centres of the magnets are. Therefore, the slowing length of such a slower is shorter than the physical size of the slower and the magnets are not used efficiently. 

The design of a more efficient longitudinal Zeeman slower is shown in fig.\,5. This slower consists of four identical 
sections, each of which includes two stacks of magnets (``entrance'' and ``output'' magnets) of opposite polarity, placed symmetrically with respect to the $z$-axis and located in $X0Z$ and $Y0Z$ planes. Each ``entrance'' magnet consists of 16 individual neodymium 
magnets with length of 8\,mm, diameter of 20\,mm and magnetic moment of $M=2.2$\,A$\cdot$m$^2$. Each ``output'' magnet contains 6 individual magnets.
The exact distances of the centres of these magnets from the $z$-axis, which are changing nearly linearly with $z$-coordinate, are given in Table\,1.

Unfortunately, the problem of finding of the positions of individual magnetic dipoles of the slower, which is producing the 
target spatial distribution of the field of the slower at its axis, has no single solution. The exact fitting of the target field distribution at the ends of the slower is also impossible.

Our general procedure for finding the positions of the magnets consists of the following steps. First, a simple spin-flip slower with two long cylindrical magnets, placed parallel to the $z$-axis as in fig\,3, is calculated to have its field distribution maximally close to the target field distribution. As a next step, the magnets are moved closer to the axis of the slower, in such a way that the corresponding distances of the magnets $x_i$ from the axis are linearly decreasing from the centre of the slower (where the magnetic field crosses zero) towards the outer ends of the slower. Finally, the residual kinks in the spatial distribution of the magnetic field are eliminated by fine tuning of the positions of the group of the magnets in the vicinity of that kink.

\begin{figure}
\includegraphics[scale=0.4]{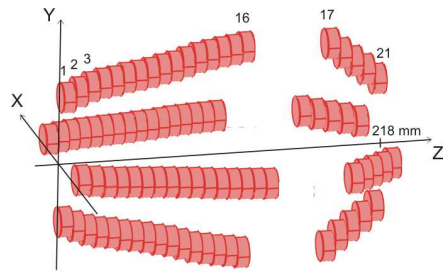}
\caption{Design of the Sr longitudinal MD Zeeman slower. Each cylindrical magnet is 8\,mm long and 20\,mm in diameter.}
\end{figure}

This slower was designed for $\epsilon=0.92$, capture velocity $v_c=410$\,m/s, final velocity $v_f=25$\,m/s, saturation parameter at the end of the slower $s_0=1.0$ and diameter of the cooling laser beam at that location of 1.0\,cm with a convergence angle of $\theta=0.014$\,rad. The design frequency detuning of the cooling laser beam is $\delta_0/2\pi=476$\,MHz.
The spatial distribution of the magnetic field of the slower, calculated with the MD model,
and the target field distribution are shown in fig\,6.
The slowing length is now $18$\,cm, which is close to the physical length of the slower.

\begin{figure}
\includegraphics[scale=0.4]{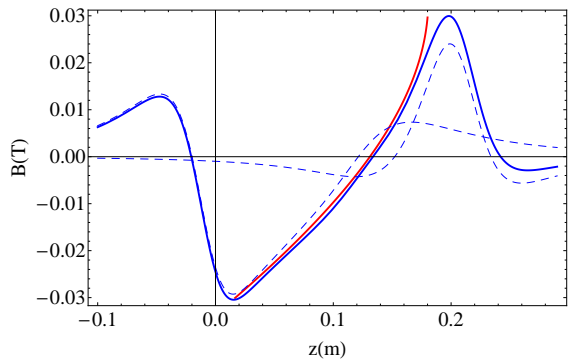}
\caption{Spatial distribution of the magnetic field along the axis of the Sr longitudinal MD Zeeman slower.}
\end{figure}

\begin{figure}
\includegraphics[scale=0.4]{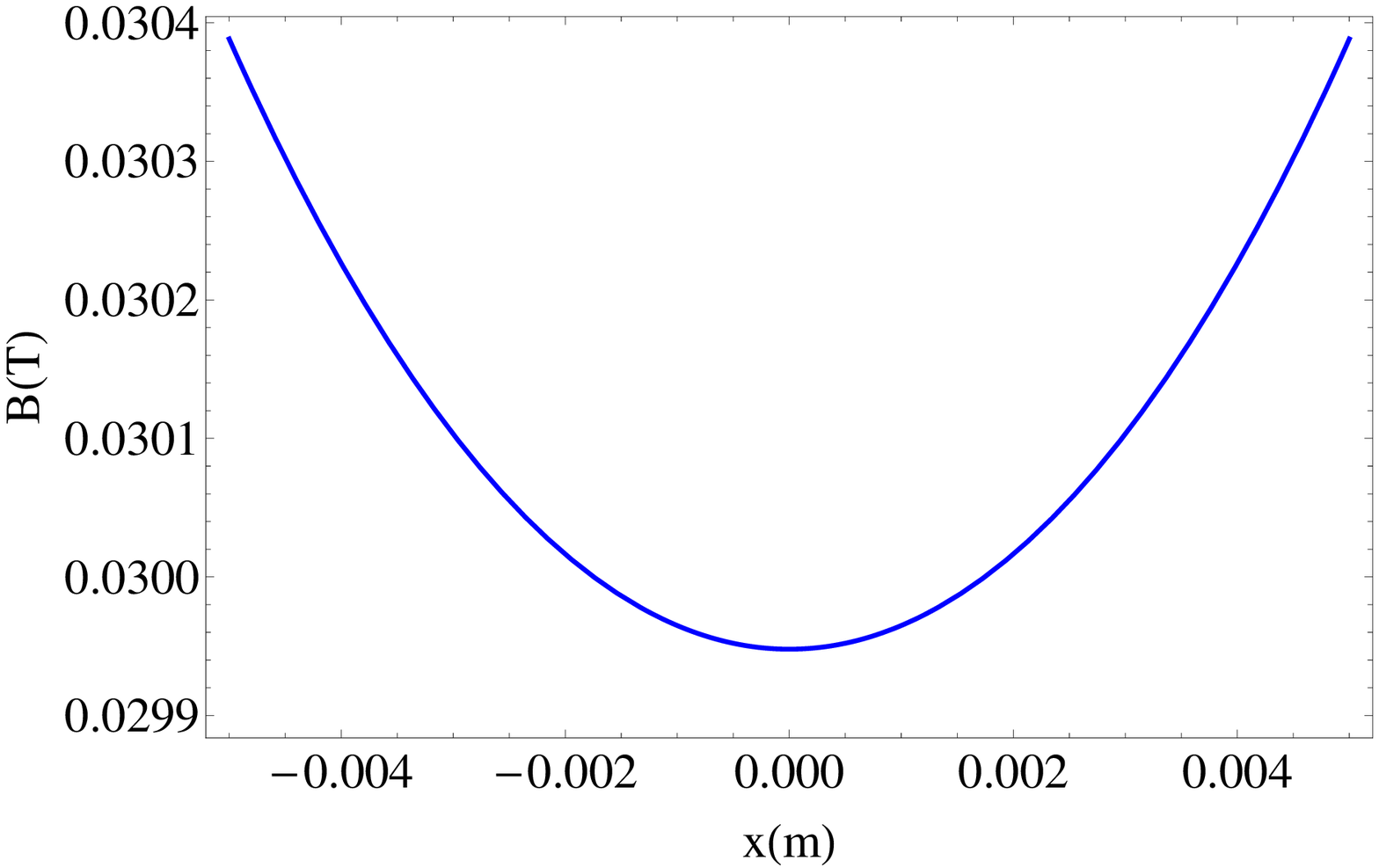}
\caption{Spatial distribution of the magnetic field flux along the x-axis at the output of the Sr longitudinal MD Zeeman slower.}
\end{figure}

\begin{table}
\caption{Calculated positions of the magnets in the Sr slower.}
\label{tab:1}
\begin{tabular}{lllllllllllllllll}
\hline\noalign{\smallskip}
$i$ & 1 & 2 & 3 & 4 & 5 & 6 & 7 & 8 & 9 & 10 & 11 & 12 & 13 & 14\\
\noalign{\smallskip}\hline\noalign{\smallskip}
$z_i$(mm) & 0 & 8 & 16 & 24 & 32 & 40 & 48 & 56 & 64 & 72 & 80 & 88 & 96 & 104 \\
$x_i$(mm) & 43 & 47 & 50 & 52 & 53 & 55 & 56 & 57 & 59 & 61 & 63 & 66 & 67 & 69 \\
\noalign{\smallskip}\hline\hline
$i$ & 15 & 16 & 17 & 18 & 19 & 20 & 21\\
\noalign{\smallskip}\hline\noalign{\smallskip}
$z_i$(mm)  & 112 & 120 & 178 & 186 & 194 & 202 & 210\\
$x_i$(mm) & 70 & 71 & 70 & 63 & 57 & 51 & 44\\
\noalign{\smallskip}\hline
\end{tabular}
\end{table}

To calculate the magnetic field distribution of the MD-Zeeman
slower in the $xy$-plane one needs to take into account all
three components of the magnetic field $B_x$, $B_y$ and $B_z$,
which define the total amplitude of the magnetic field
$\left| B(x,y,z) \right|=\sqrt{B_x^2+B_y^2+B_z^2}$ responsible for the local
Zeeman shift of atomic magnetic states. The transverse variation
of the field is the strongest at the input and output edges of
the Zeeman slower, where the distances of the magnets from the axis of
the slower are the smallest. The calculations show that within a region
with diameter of 1\,cm the field is changing by not more than 1.5\% (see fig.\,7).
The transverse component of the magnetic field at the end of
the slower and at $x=\pm 0.5$\,cm and  $y=\pm 0.5$\,cm is only $\pm0.8$\,G.
In the central region of the slower, where the distances of the magnets from the
$z$-axes of the slower are larger, the transverse variation of
the field is much smaller.

In addition, a precise calculation of the MD-Zeeman slower based on
finite-size neodymium magnets with diameter $20\,$mm and height $8\,$mm was performed with the finite
element analysis software COMSOL. These calculations show that the
optimal distances of the centres of the magnets from the
axis of the slower are exactly the same as in the MD model.

\section{Longitudinal MD-slower for Rb atoms}
\label{sec:3}

Rubidium is a popular and convenient element for laser cooling, which is used in multiple atomic physics experiments, including producing BEC and laser-cooled atomic frequency standards. Therefore, there are many traditional Zeeman slowers of different configurations that have been implemented in different labs. 
The Rb Zeeman slowers use the $\left|5S_{1/2},F=2\right> \rightarrow \left|5P_{3/2},F'=3\right>$ optical transition, which has a wavelength of $\lambda=780$\,nm and a linewidth, $\Gamma/2\pi=6.07$\,MHz. The strength of this transition and the photon recoil momentum of the Rb atom are smaller than for Sr, which makes the typical length of a Rb Zeeman slower equal to 1\,m.  
Another problem of Zeeman slowing $^{87}$Rb atoms is the crossing between the $\left|5P_{3/2},F'=3,m=-3\right>$ and $\left|5P_{3/2},F'=2,m=-1\right>$ magnetic sublevels of the excited $5P_{3/2}$ state, which takes place at $B=120$\,G. 
For the case of imperfect circular polarization of the cooling laser beam, the $\left|5P_{3/2},F'=2,m=-1\right>$ state can be excited, which can decay into the ``dark'' ground hyperfine state $\left|5S_{1/2},F=1\right>$.
The atoms in the $\left|5S_{1/2},F=1\right>$ state are out of resonance with the cooling light and therefore are lost
from further deceleration. To preserve that from happening, Rb Zeeman slowers are usually is designed in such a way that 
they include an additional uniform offset field of about 200\,G \cite{Streed,Cheiney}.

On the other hand, in the work \cite{Slowe} the spin-flip longitudinal Zeeman slower has a field that changes from $200$\,G to $-100$\,G. This slower is using an additional repumping light beam, which is tuned close to the $\left|5S_{1/2},F=1\right> \rightarrow \left|5P_{1/2},F'=2\right>$ transition on the $D_1$ manifold and has the same frequency detuning as the main cooling laser beam. In this setup a record flux of slowed atoms of $3.2\times10^{12}$\,at/s has been demonstrated.

Our design of the spin-flip longitudinal Zeeman slower for $^{87}$Rb atoms, based on permanent magnets, is shown in fig\,8.
The magnetic field of the slower is changing from -118\,G to 116\,G within its length of 0.56\,m.
Each of the magnets of the slower has a diameter 30\,mm, height 20\,mm and magnetic moment $M=12.4$\,A$\cdot$m$^2$, while some of them are combined in longer cylindrical magnets. The exact positions of the centres of these magnets are given in Table\,2.

The slower is designed for $\epsilon=0.8$, capture velocity $v_c=275$\,m/s and final velocity $v_f=25$\,m/s. It is assumed that the cooling beam has a diameter at the end of the slower of 2.0\,cm, where the saturation parameter is $s_0=3.0$, and has a convergence angle of $\theta=0.02$\,rad.
The design frequency detuning of the cooling laser beam is $\delta_0/2\pi=196$\,MHz.
The spatial distribution of the magnetic field of the slower along its axis, calculated with the COMSOL software, is shown in fig\,9. 
The sharp steps in the computed field are not real and are related to the errors due to the relatively large size of the mesh, which was used
in the tree-dimensional finite-element computations.  The last fact is confirmed by the MD-model simulations, which are in good agreement with the COMSOL results.
The uniformity of the field in the transverse direction is very good. For example, at the output of the slower at $z=0.57$\,m the increase in field at $x=\pm0.5$\,cm is about 0.3\,G only.

One great advantage of the given spin-flip Zeeman slower for Rb is that its maximal Zeeman shift is only about 160\,MHz.
This makes it easy to generate the proper frequencies for the MOT and Zeeman slower cooling laser beams with standard single-pass acousto-optical modulators.

\begin{table}
\caption{Calculated positions of the magnets in the Rb slower.}
\label{tab:2}
\begin{tabular}{lllllllllllllllllll}
\hline\noalign{\smallskip}
$i$ & 1 & 2 & 3 & 4 & 5 & 6 & 7 & 8 & 9 & 10\\
\noalign{\smallskip}\hline\noalign{\smallskip}
$z_i$(mm) & 10 & 30 & 50 & 70 & 90 & 110 & 130 & 150 & 170 & 190 \\
$x_i$(mm) & 106 & 111 & 115 & 119 & 123 & 126 & 127 & 129 & 133 & 138 \\
\noalign{\smallskip}\hline\hline
$i$ & 11 & 12 & 13 & 14 & 15 & 16 & 17 & 18 & 19\\
\noalign{\smallskip}\hline\noalign{\smallskip}
$z_i$(mm)  & 210 & 230 & 250 & 270 & 290 & 310 & 330 & 350 & 370\\
$x_i$(mm) & 142 & 147 & 153 & 159 & 166 & 172 & 177 & 181 & 185\\
\noalign{\smallskip}\hline
\noalign{\smallskip}\hline\hline
$i$ & 20 & 21 & 22 & 23 & 24 & 25\\
\noalign{\smallskip}\hline\noalign{\smallskip}
$z_i$(mm)  & 495 & 515 & 535 & 555 & 576 & 595\\
$x_i$(mm) & 184 & 170 & 157 & 141 & 123 & 102\\
\noalign{\smallskip}\hline
\end{tabular}
\end{table}

\begin{figure}
\includegraphics[scale=1.0]{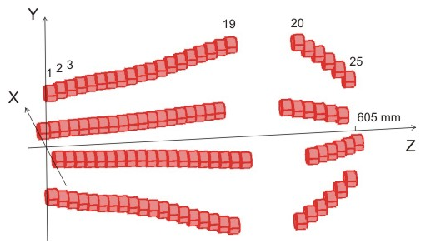}
\caption{Design of the Rb longitudinal MD Zeeman slower. Each cylindrical magnet is 20\,mm long and 30\,mm in diameter.}
\end{figure}

\begin{figure}
\includegraphics[scale=0.4]{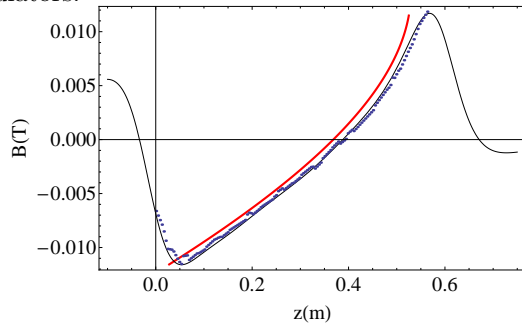}
\caption{Thick solid line - the target spatial distribution of the magnetic field of the Rb longitudinal Zeeman slower, calculated for $\epsilon=0.8$, capture velocity of $v_c=275$\,m/s and final velocity $v_f=25$\,m/s. The dotted line shows the corresponding field distribution calculated with COMSOL software. The thin solid line is the field distribution calculated with MD model.}
\end{figure}

\section{Conclusion}
\label{sec:4}

Examples of longitudinal spin-flip Zeeman slowers for Sr and Rb atoms, composed of arrays of
compact permanent magnets, are presented. 
The advantage of these longitudinal slowers, compared to the transverse MD-Zeeman slowers, is that they
demand only half as much power for the cooling laser light, which is important for some atoms, like Sr or Yb.
In addition, the longitudinal spin-flip slowers work for all laser-cooled atoms, which is not the case for the
transverse slowers.
On the other hand, the longitudinal MD-Zeeman slowers are more complicated, compared to the transverse ones, and 
demand larger magnets.   
The slowers do not consume electrical energy, and the spatial distribution of their magnetic field
and the total length of the slowers are easily adjustable. The calculated spatial distribution of the magnetic field of the slower is extremely smooth, which should make it possible to operate it at the optimal conditions, where the length of the slower is minimal, which reduces the transverse expansion of the
slowed atoms.

\section*{Acknowledgment}

Many thanks to Rachel Godun for the valuable comments.
This work was funded by the UK National Measurement System Directorate of the Department of Trade and Industry.

\end{document}